\documentstyle[prl,aps,multicol,psfig]{revtex}
\begin{document}

\draft

\title{Explosion of a Collapsing Bose-Einstein Condensate}

\author{R.A. Duine and H.T.C. Stoof}
\address{Institute for Theoretical Physics,
         University of Utrecht, Princetonplein 5, \\
         3584 CC Utrecht, The Netherlands}

\maketitle

\begin{abstract}
We show that elastic collisions between atoms in a Bose-Einstein
condensate with attractive interactions lead to an explosion that
ejects a large fraction of the collapsing condensate. We study
variationally the dynamics of this explosion and find excellent
agreement with recent experiments on magnetically trapped
$^{85}$Rb. We also determine the energy and angular distribution
of the ejected atoms during the collapse.
\end{abstract}


\begin{multicols}{2}
{\it Introduction.} --- Ever since the achievement of
Bose-Einstein condensation in an atomic gas with attractive
interactions \cite{curtis}, it has been an important objective to
study in detail the condensate collapse that is predicted to occur
\cite{collapse1,collapse2} if the number of condensate atoms
exceeds a certain maximum number determined by the strength of the
attractive interactions \cite{keith}. In the pioneering
experiments of Bradley {\it et al}.\ this objective was very
difficult to achieve for two reasons. First, the use of doubly
spin-polarized atomic $^7$Li results for these experiments in a
maximum number of condensate atoms of only about 1400 atoms, which
is too small to allow for nondestructive imaging of the collapse
dynamics. Second, the inherent stochastic nature of the growth and
collapse cycles that occur because one aims at evaporatively
cooling far below the critical temperature \cite{cass1}, prevents
sufficient control over the initial conditions of the condensate
to perform a sequence of destructive measurements. However, a
statistical analysis of the condensate growth and collapse cycles
has nevertheless revealed important information on the collapse
process \cite{cass2}, and new experimental techniques are
presently being applied to overcome in particular the latter of
these problems \cite{randy}.

Complementary to these developments, Cornish {\it et al}.\ have
recently been able to overcome both problems in an ingenious
experiment with spin-polarized atomic $^{85}$Rb \cite{simon}. In
this experiment one makes use of the fact that the $f=2$, $m_f=-2$
hyperfine ground state of $^{85}$Rb has a so-called Feshbach
resonance \cite{eite} at a magnetic field of about $1.55$ mT,
which offers the opportunity to magnetically tune the interatomic
interactions from strongly repulsive to very attractive. As a
consequence Cornish {\it et al}.\ were able to first produce a
large, stable and essentially pure condensate, and then suddenly
switch the interactions from repulsive to attractive to induce a
collapse and observe its properties.

Perhaps the most surprising outcome of this experiment is that
during the collapse an explosion occurs that ejects a large number
of relatively highly energetic atoms from the condensate. A
possible mechanism that immediately comes to mind for the
production of these energetic atoms are inelastic collisions
between condensate atoms that flip a spin and, therefore, convert
Zeeman energy into kinetic energy. However, a simple estimate
shows that this mechanism generally leads to particles with much
too high energies and is also incapable of ejecting so many atoms
from the condensate. Similarly it appears that elastic collisions
between a condensate atom and an atom in the thermal cloud, which
in principle cannot experimentally be excluded to be present, also
occur too infrequent to explain the experimental results. We are
thus faced with the important theoretical task to determine what
physical mechanism is responsible for the observed explosion. It
is the main purpose of this Letter to point out that elastic
collisions between two condensate atoms can provide such a
mechanism.

{\it Elastic Condensate Collisions.} --- In detail, the scattering
process that we have in mind is a collision between two condensate
atoms by which one of the atoms is stimulated back into the
condensate again but the other atom is ejected out of the
condensate. It is important to realize that such a process is
forbidden for a homogeneous condensate due to momentum
conservation. However, for a trapped gas this is no longer true,
because the condensate now occupies a band of low-lying momentum
states. Note also that momentum and energy conservation ensure
that the ejected atoms will automatically have, as compared to
inelastic spin-flip collisions, small kinetic energies and can
thus remain trapped as seen in experiment. To see if the above
mechanism can also explain the large amount of ejected atoms, we
need to calculate the rate associated with this process.

This can be most easily achieved by using the fact that we are
dealing with a two-body interaction process. As a result the rate
of change of the total number of condensate atoms $N_{\rm c}(t)$ is
related to the instantaneous interaction energy of the condensate
$E^{\rm int}_{\rm c}(t)$ by
\begin{equation}
\label{rate}
\frac{d N_{\rm c}(t)}{dt} =  \frac{2}{\hbar} {\rm Im}
                                     \big[ E^{\rm int}_{\rm c}(t) \big]~.
\end{equation}
In the Bogoliubov theory of the dilute Bose gas we simply have
that
\begin{equation}
E^{\rm int}_{\rm c}(t) = \frac{T^{\rm 2B}}{2} \int d{\bf x}~
                                          |\phi({\bf x},t)|^4~,
\end{equation}
which in momentum space reads
\begin{eqnarray}
\label{energy}
E^{\rm int}_{\rm c}(t) = \frac{1}{2}
  \left( \prod_{i=1}^4 \int \frac{d{\bf k}_i}{(2\pi)^3} \right)~
     \phi^*({\bf k}_4,t)  \phi^*({\bf k}_3,t)  \hspace*{0.3in} \nonumber \\
     \times (2\pi)^3
         \delta({\bf k}_4 + {\bf k}_3 - {\bf k}_2 - {\bf k}_1) T^{\rm 2B}
         \phi({\bf k}_2,t) \phi({\bf k}_1,t)~.
\end{eqnarray}
Here the two-body T(ransition) matrix equals $T^{\rm 2B} = 4\pi
a\hbar^2/m$ in terms of the $s$-wave scattering length $a$ and the
mass $m$ of the $^{85}$Rb atoms, and $\phi({\bf x},t)$ is the
condensate wave function. In this approximation the condensate
interaction energy is thus purely real and the number of
condensate atoms is conserved.

However, a microscopic derivation \cite{henk,allan} shows that in
Eq.~(\ref{energy}) the expression $(2\pi)^3 \delta({\bf k}_4 +
{\bf k}_3 - {\bf k}_2 - {\bf k}_1) T^{\rm 2B}$ should in principle be
replaced by the many-body T-matrix element $T^{\rm MB}({\bf k}_4, {\bf
k}_3, {\bf k}_2, {\bf k}_1)$, whose real part is at low
temperatures indeed well approximated by the Bogoliubov result
$(2\pi)^3 \delta({\bf k}_4 + {\bf k}_3 - {\bf k}_2 - {\bf k}_1)
T^{\rm 2B}$. However, the many-body T matrix describes also real
(incoherent) collisions taking place in the gas and as a result
acquires an imaginary part that can be obtained from an optical
theorem \cite{henk}. Being interested in the ejection of relatively highly
energetic particles, we can in first instance neglect the
mean-field effects on the corresponding intermediate state to
obtain
\end{multicols}
\begin{eqnarray}
\label{imag}
{\rm Im}
 \big[ T^{\rm MB}({\bf k}_4, {\bf k}_3, {\bf k}_2, {\bf k}_1) \big]
                                     \hspace*{5.0in} \nonumber \\
   = - 2\pi (T^{\rm 2B})^2 \int \frac{d{\bf k}}{(2\pi)^3}~
        \delta(\epsilon({\bf k}_2) + \epsilon({\bf k}_1)
               - \epsilon({\bf k})
               - \epsilon({\bf k}_2 + {\bf k}_1 - {\bf k}))~
         \phi({\bf k}_4 + {\bf k}_3 - {\bf k},t)
         \phi^*({\bf k}_2 + {\bf k}_1 - {\bf k},t)~,
\end{eqnarray}
\begin{multicols}{2}
\noindent
where $\epsilon({\bf k}_i) = \hbar^2 {\bf k}_i^2/2m$,
$\hbar {\bf k}$ is the momentum of the ejected atom, and the
dependence on the condensate wave function is caused by the effect
of Bose enhancement of scattering into already occupied states.
Interestingly, the latter is here nondiagonal in momentum space
due to the inhomogeneity of the condensate.

\begin{figure}
\psfig{figure=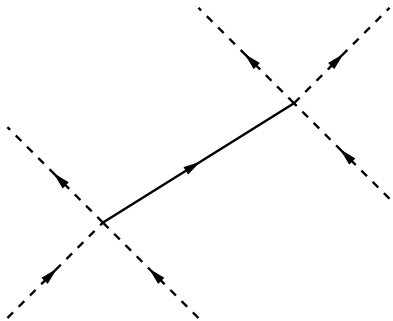}
\caption{\narrowtext
         The imaginary part of this Feynman diagram determines the rate for
         ejection of atoms out of the condensate due to elastic condensate
         collisions. The dashed lines correspond to condensate atoms and the
         full line to a noncondensate atom.
         \label{fig1}}
\end{figure}

Notice that from a field-theoretical point of view, we have in the
above manner just arrived at an evaluation of the Feynman diagram
drawn in Fig.~\ref{fig1}. We thus conclude that we are effectively
dealing with a three-body process. Indeed, by taking the
functional derivative $\delta/\delta \phi^*({\bf x},t)$ of our
result for the condensate interaction energy, we can obtain an imaginary
three-body correction term to the Gross-Pitaevskii equation of the
condensate, which provides another way of deriving the desired
ejection rate. It is also worth mentioning that we qualitatively expect
mean-field effects to reduce the above ejection rate in the case of a gas
with positive scattering length, because it then costs energy to
remove a particle from the condensate. Such an effect, however, does
not occur for the negative scattering length case of interest here.

{\it Collapse Dynamics.} --- To make further progress we need to
determine the condensate wave function during the collapse. We do
this by solving the Gross-Pitaevskii equation variationally,
taking for the wave function an anisotropic Gaussian with
time-dependent widths denoted by $q_x(t)$, $q_y(t)$, and $q_z(t)$.
In detail we use, apart from a for our purposes irrelevant phase
factor,
\begin{eqnarray}
\phi({\bf x},t) = \sqrt{N_{\rm c}}
  \prod_j \left( \frac{1}{\pi q_j^2(t)} \right)^{1/4}
                                            \hspace*{1.0in} \nonumber \\
  \times \exp \left\{ - \frac{x_j^2}{2q_j^2(t)}
                      \left( 1 - i \frac{m q_j(t)}{\hbar}
                                       \frac{d q_j(t)}{dt} \right)
         \right\}~.
\end{eqnarray}
The reason for this Gaussian {\it ansatz} is that it is the
appropriate description of the condensate after the explosion has
occurred and the remnant contains only a relatively small amount
of atoms \cite{collapse2,tony}. Moreover, it is well-known that a
Gaussian {\it ansatz} gives excellent results for the collective
modes of the condensate, even when the number of atoms in the
condensate is so large that the Gaussian approximation for the
condensate wave function is strictly speaking no longer valid
\cite{gauss}. We thus expect it to give physically sensible
results for the dynamics of the collapse at all times.

Within this approach the explosion of the collapsing condensate is described by
the `classical' equations of motion
\begin{equation}
m \frac{d^2 q_j(t)}{dt^2}
       = - \frac{\partial}{\partial q_j} V({\bf q}(t);N_{\rm c}(t))~,
\end{equation}
with a potential energy that is equal to
\begin{equation}
V({\bf q};N_{\rm c}) =
 \sum_j \left( \frac{\hbar^2}{2m q_j^2} + \frac{m \omega_j^2 q_j^2}{2} \right)
 + \sqrt{\frac{2}{\pi}} \frac{a \hbar^2 N_{\rm c}}{m q_x q_y q_z}
\end{equation}
for an anisotropic harmonic trapping potential with spring
constants $m \omega_j^2$. Moreover, these equations are coupled to
the rate equation for the number of condensate atoms that follows
from Eqs.~(\ref{rate}) and (\ref{imag}). In the anisotropic case
this cannot be worked out fully analytically and we have to
determine the ejection rate numerically. However, in the isotropic
case we simply find that
\begin{equation}
\label{iso}
\frac{d N_{\rm c}(t)}{dt} = - \frac{2 \sqrt{5}}{\pi}
                              \frac{a^2 \hbar N_{\rm c}^3(t)}{m q^4(t)}~.
\end{equation}

The dependence of the right-hand side on $N_{\rm c}(t)$ and $q(t)$
is somewhat unusual, because if we had discussed the effect of
inelastic two-body processes the rate of change of the number of
condensate atoms would be proportional to $N_{\rm c}^2(t)/q^3(t)$.
The additional factor of $N_{\rm c}(t)$ is easily understood and
reflects the fact that the ejection of the atoms is Bose
stimulated. As mentioned previously, our mechanism therefore
effectively behaves as a three-body process. The additional factor
of $1/q(t)$ is more subtle and shows that, if the condensate
collapses and the wave function becomes more spread out in
momentum space, more atoms can satisfy the energy and momentum
constraints for the ejection. Having said this, we should make
sure that the dominant contribution to the integration in the
right-hand side of Eq.~(\ref{imag}) comes from momenta $\hbar {\bf
k}$ that lie outside the band of momenta occupied by the
condensate. We can most conveniently achieve this by including in
the integrand the factor $(1 - |\phi({\bf k},t)|^2/|\phi({\bf
0},t)|^2)$, which smoothly interpolates between $0$ for ${\bf k}
\cdot {\bf q} \ll 1$ and $1$ for ${\bf k} \cdot {\bf q} \gg 1$. In
the following we always use this smooth cut-off, because atoms
with momenta below the cut-off are most likely stimulated back
into the condensate and do not contribute to the ejection rate. It
has, however, only a small quantitative effect as can be seen
explicitly in the isotropic case, where the right-hand \linebreak

\begin{figure}
\psfig{figure=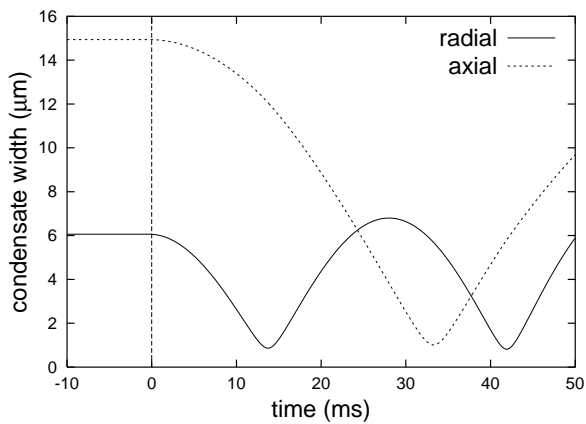}
\caption{\narrowtext
         The radial and axial widths of the condensate during a collapse
         induced by a sudden change of sign in the scattering length. At the
         origin of the time axis the scattering length vanishes. See
         text for more details.
         \label{fig2}}
\end{figure}

\begin{figure}
\psfig{figure=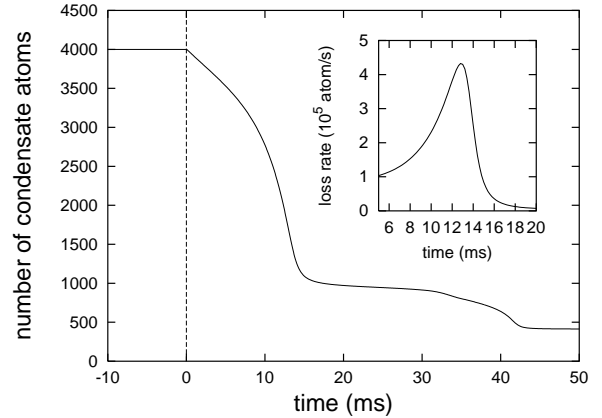} \caption{\narrowtext
         The number of atoms in the condensate during a collapse
         induced by a sudden change of sign in the scattering length. The
         inset shows the ejection rate.
         \label{fig3}}
\end{figure}

\noindent side of Eq.~(\ref{iso}) just becomes multiplied by the
factor $(1-(8/25)\sqrt{2/5}) \simeq 0.8$.

We have solved the above set of coupled equations for the
conditions of the experiment performed by Cornish {\it et al.}
\cite{simon}. In this experiment one uses a cigar-shaped magnetic
trap with a radial frequency of $\omega_r = \omega_x = \omega_y =
2\pi \times 17.5$ Hz and an axial frequency of $\omega_z = 2\pi
\times 6.8$ Hz. Moreover, one first makes a large condensate with
about $4000$ atoms and a positive scattering length of $2500$
$a_0$. The scattering length is then within $0.5$ ms changed to $-
60$ $a_0$ by means of a linear ramp in the magnetic bias field.
The outcome of our simulation of this experiment is summarized in
Figs.~\ref{fig2} and \ref{fig3}, and appears to be in excellent
agreement with the preliminary experimental data \cite{carl}. From
these figures we see that the condensate collapses first in the
radial direction in approximately $\pi/(2\omega_r) \simeq 14$ ms.
During the last part of this collapse an explosion occurs in which
about $3/4$ of the initial number of atoms is expelled from the
condensate. As a result the number of condensate atoms is now less
than the maximum number of atoms possible to have a metastable
condensate, and the radial collapse of the condensate is turned
into a large amplitude oscillation. In principle we can discern at
each inner turning point of the radial and axial oscillation an
increased loss of condensate atoms, but these are never as
dramatic as in the first time, since the condensate is now no
longer unstable.

\begin{figure}
\psfig{figure=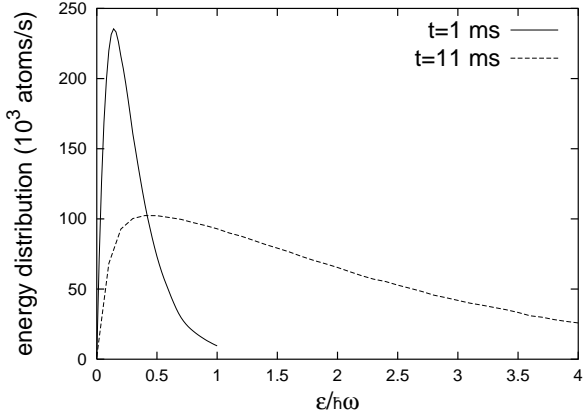}
\caption{\narrowtext
         The energy distribution of the atoms ejected from the condensate
         at two times during the first radial collapse. We have used a
         dimensionless energy variable by scaling the energy with
         $\hbar \omega = \hbar (\omega_r^2 \omega_z)^{1/3} \simeq 0.61$ nK.
         \label{fig4}}
\end{figure}

{\it Explosion Process.} --- To explore the physics of the
explosion further, we have also determined the energy and angular
distribution of the ejected atoms, which have not yet been
examined in detail experimentally. In Fig.~\ref{fig4} we show the
energy distribution at two times during the first radial collapse
of the condensate. What is most striking is that at later times
the distribution is much broader. This is again caused by the
condensate wave function being much more spread out in momentum
space at the

\begin{figure}
\psfig{figure=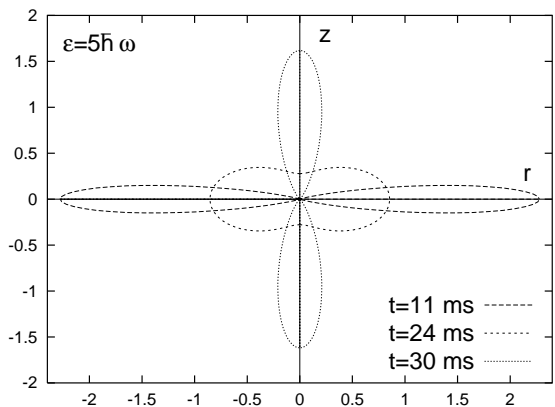}
\caption{\narrowtext
         Polar plot of the angular probability distribution of atoms ejected
         from the condensate at three different times but at the same energy.
         \label{fig5}}
\end{figure}

\noindent later time, so that the ejected atoms can gain much more
energy from the condensate collisions. In Fig.~\ref{fig5} we have
also depicted at various times the angular probability
distribution, which turns out to be essentially independent of the
energy of the ejected atoms but is clearly seen to depend strongly
on the ratio of the radial and axial widths of the condensate. In
particular, at the peak of the explosion the distribution is very
anisotropic and almost no atoms are being ejected along the
z-axis. This can be understood from the fact that the condensate
has at that time a very elongated cigar shape. As a result both
the total momentum of the two colliding condensate atoms as well
as the momentum of the atom that is stimulated back into the
condensate, always have to be directed almost perpendicular to the
z-axis. The same is therefore true for the momentum of the ejected
atom.

{\it Discussion.} --- Although we have focused in this Letter on
the recent exciting experiments with $^{85}$Rb, we believe that
the new condensate loss mechanism that we have proposed here is
also important for the experiments with $^7$Li and may resolve the
existing discrepancy between theory and experiment in that case
\cite{cass2}. In this respect it should be noted that the collapse
occurring in the $^{85}$Rb experiments is physically quite
different and in a sense not so violent than the one occurring in
$^7$Li. Roughly speaking the difference is that in the former case
the large condensate ejects so many particles that the remnant of
the explosion corresponds to a metastable condensate, whereas in
the latter case it is precisely this metastable condensate that
collapses to even smaller sizes. Possibly another important
difference is that in the experiments of Cornish {\it et al.} no
thermal component is visible, whereas the experiments of Sackett
{\it et al.} are close to the critical temperature and, therefore,
a large thermal cloud is constantly feeding the condensate. We
intend to come back to a detailed theoretical treatment of these
interesting issues in a future publication.

It is our pleasure to thank Simon Cornish and Carl Wieman for
providing us with their experimental data before publication and
for stimulating discussions that have initiated the above. This
work is supported by the Stichting Fundamenteel Onderzoek der
Materie (FOM), which is financially supported by the Nederlandse
Organisatie voor Wetenschappelijk Onderzoek (NWO).

\end{multicols}
\end{document}